\documentclass[twocolumn,prl,nobalancelastpage,amssymb,superscriptaddress,showpacs]{revtex4-1}

\usepackage{Symbols}
\usepackage{Shortcuts}
\usepackage{Text}

\usepackage{epsfig}
\usepackage{subfigure}
\usepackage{slashbox}
\usepackage[colorlinks=true,citecolor=blue,linkcolor=blue]{hyperref}
\usepackage{pdfpages}
\makeatletter
\AtBeginDocument{\let\LS@rot\@undefined}
\makeatother

\begin{document}
\title{Detection of quantum interference without interference}

\date{\today}

\author{Iliya Esin}
\affiliation{\mbox{Physics Department, Technion, 3200003 Haifa, Israel}}
\author{Alessandro Romito}
\affiliation{\mbox{Department of Physics, Lancaster University, Lancaster LA1 4YB, United Kingdom}}
\author{Yuval Gefen}
\affiliation{\mbox{Department of Condensed Matter Physics, The Weizmann Institute of Science, Rehovot 76100, Israel}}

\begin{abstract}
Quantum interference is typically detected through the dependence of the interference signal on certain parameters (path length, Aharonov-Bohm flux, etc.), which can be varied in a controlled manner.  The destruction of interference by a which-path measurement is a paradigmatic manifestation of quantum effects.  Here we report on a novel measurement protocol that realizes two objectives: (i)  certifying that a measured signal is the result of interference avoiding the need to vary parameters of the underlying interferometer, and (ii) certifying that the interference signal at hand is of quantum nature. In particular, it yields a null outcome in the case of classical interference. Our protocol comprises measurements of cross-correlations between the readings of which-path weakly coupled detectors positioned at the respective interferometer's arms and the current in one of the interferometer's drains. We discuss its implementation with an experimentally available platform: an electronic Mach-Zehnder interferometer (MZI) coupled electrostatically to ``detectors'' (quantum point contacts).
\end{abstract}
\maketitle

\paragraph{Introduction ---}

Quantum interferometry differs from  its classical counterpart in its sensitivity to ``which-path'' detection.
In classical wave interference, the wave amplitude can be observed along individual interfering trajectories without affecting the interference itself. Quantum mechanically, information on the trajectory traveled by the interfering particle destroys the interference pattern.
This is a specific example of the adverse effect of quantum measurement: it is an invasive operation, accompanied by back-action of the detector on the system's state \cite{Wiseman2009,Jacobs2014} and, in the case of strong (projective) measurement, it leads  to the collapse of the system's wave function \cite{Neumann1955}. As far as establishing the fact that interference, classical or quantum, takes place, common wisdom is that this requires continuous variation of a control parameter (e.g, interferometer arm’s length, Aharonov Bohm (AB) flux for charged particles \footnote{In principle one can assert the quantum nature of interference if the latter depends on an Aharonov Bohm (AB) flux, which is a quantum effect per se. However, in practically all instances of solid-state implementations of AB interferometry, direct magnetic field, as opposed to AB flux, is applied}).
The observation of interference {\it and} of the collapse of the coherent wavefunction to a state that does not exhibit an interference pattern (following which-path detection) are a manifestation of the quantum nature of the phenomenon.
Such combined measurements have been demonstrated in studies of average currents of electronic interferometers \cite{Yacoby1994,Buks1998,Neder2007,Dressel2012,Weisz2014}, and analyzed theoretically for single-electron \cite{Elitzur1993,Shpitalnik2008} and many-body \cite{Esin2016,Zilberberg2016} protocols.
The question addressed here is of fundamental nature: can one detect particle interference avoiding the need to vary an external parameter, \emph{and} verify that the interference signal is inherently of a quantum nature?

\begin{figure}
  \centering
  \includegraphics[width=8.6cm]{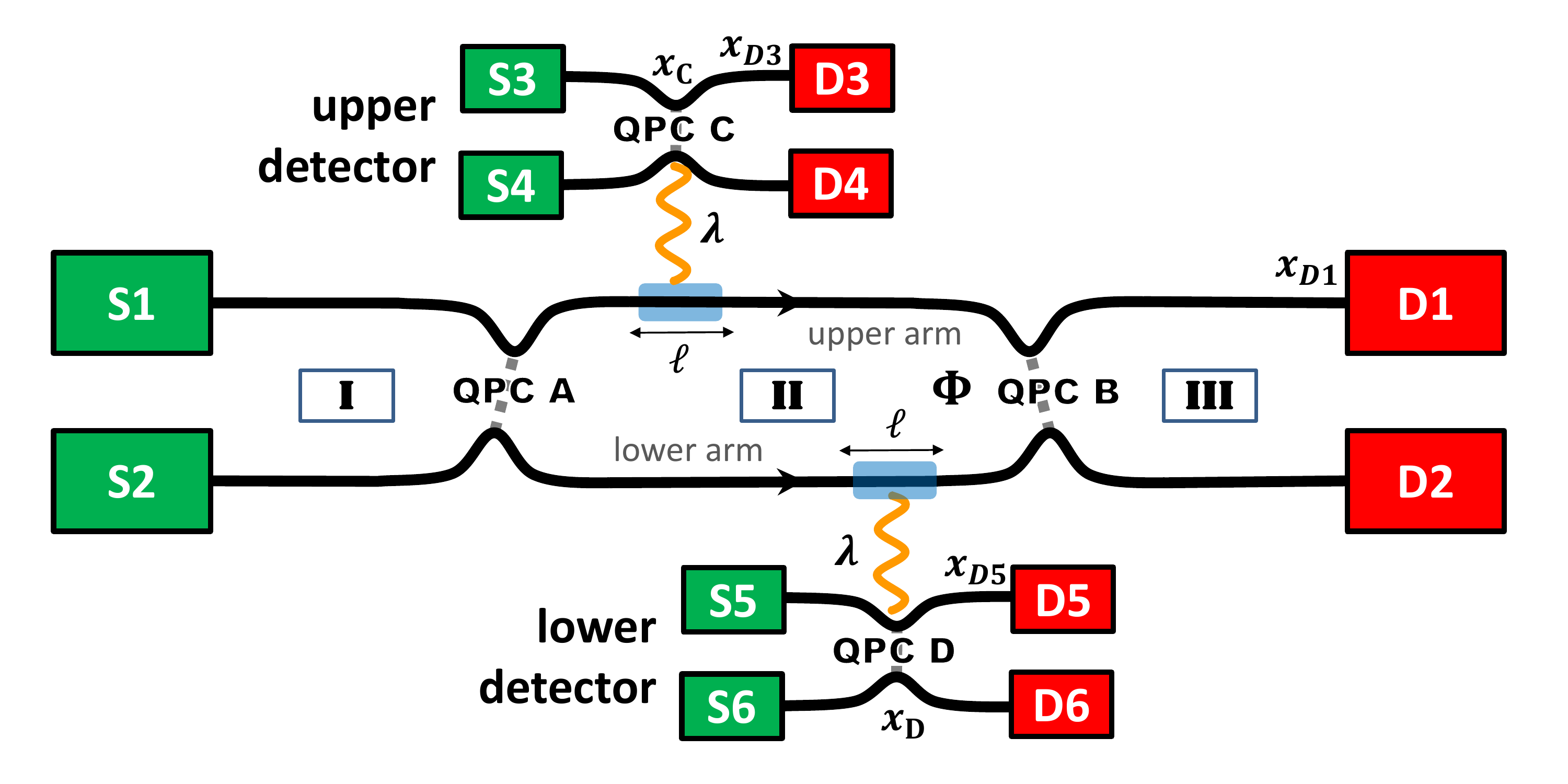}\\
  \caption{A detection setup. An MZI electrostatically coupled to two QPCs C and D serving as which-path detectors. The transmission through QPC C(D) is slightly modified (with a strength proportional to $\lm$) upon the detection of a charge fluctuation in the respective MZI's arm, within a segment $\ell$. The currents are measured in the drains $\yD1$, $\yD3$, and $\yD5$. Section II of the MZI is threaded with a magnetic flux $\F$. \label{fig:Setup}}
\end{figure}

In this work we report on a quantum measurement protocol that is used to certify the presence of quantum interference through an interferometer without varying the interferometer's parameters. We make use of minimally-invasive (weak) which-path measurements and their correlations with the interferometer signal. The non-locality of the which-path measurements provides us access to the individual wave packets that make the interference signal. Continuous (weak) measurements allow one to preserve the quantum coherence of the state since the latter is only perturbatively affected as information is being acquired by the detector \cite{Jordan2005,Clerk2010}.
Correlating the outcome of weak quantum measurements with a subsequent strong measurement forms the basis of weak values \cite{Aharonov1988,Dressel2014}; the latter has been introduced to address foundational issues \cite{Duck1989,Aharonov1990,Aharonov1991,Tollaksen2010,Aharonov2014} and, later on, for various applications \cite{Ritchie1991,Pryde2005,Romito2008,Dixon2009,Lundeen2011,Kocsis2011,Rogers2012,Bednorz2013,Romito2014,Jordan2014,Esin2016a}.
A protocol involving projective which-path measurements correlated with the input signal has been shown to violate Bell-like inequalities \cite{Santo2019}.
 Here we define and implement a more complex correlated measurement protocol involving  simultaneous (weak) detection of which-paths signals in the respective arms of the interferometer and the (strongly measured) interference signal.
Our protocol provides an experimental recipe for accessing  non-classical contributions to the interference signal. At the same time, it avoids the need for measuring interference patterns and measurement induced back-action separately.
Furthermore, to underscore the fact that our protocol addresses genuinely quantum effects, we demonstrate that when applied to a classical interferometer it yields a null outcome.
For the sake of specificity, we outline the implementation of our protocol with an electronic MZI \cite{Ji2003}.

%

\paragraph{Setup and protocol ---}

Our electronic MZI has two electronic beam splitters, (a.k.a. quantum point contacts (QPCs), cf. QPC A and B in Fig.~\ref{fig:Setup}).
The propagation of the electron through the MZI is described in terms of scattering states. For the sake of simplicity, we consider a monochromatic electron beam of energy  $\hbar \w$ that originates from the source S$n$; the respective incoming state is given by
\Eq{
	\y_{mn}(x,\w;t)=\frac{1}{\sqrt{L}} e^{-i\w\bR{t-\frac{x}{v}}} \mathcal{A}_{mn} (x,\w)  ,
	\label{eq:MZImodes}
}
where $\mathcal{A}_{mn}(x,\w)=\I_{mn} $ if $x\in \yI$, $\mathcal{A}_{mn}(x,\w)= \bS{\cS_\yA(\w)}_{mn} $ if $x\in \yI\yI$, and $\mathcal{A}_{mn}(x,\w)= \bS{\cS_\yA(\w)\cS_\yB(\w)}_{mn}$ if $x\in \yI\yI\yI$.
The sectors I, II, and III are shown in Fig.~\ref{fig:Setup} and the indices $m,n=1,2$ label the two arms of the interferometer. For a vector state $\vert \psi \rangle$ denoting the amplitudes in the two arms of the MZI, the effects of QPCs A and B are described by the scattering matrices $\cS_\yA=\mat{r_\yA&-t_\yA\as\\t_\yA&r_\yA}_{\rm MZI}\otimes \I_{\yC}\otimes \I_{\yD}$ and $\cS_\yB=\mat{r_\yB&-t_\yB\as e^{i\ch}\\t_\yB e^{-i\ch}&r_\yB}_{\rm MZI}\otimes \I_{\yC}\otimes \I_{\yD}$ respectively, where $t_\yA$ ($t_\yB$) is the amplitude for an electron incoming from arm $1$ to be transmitted to arm $2$, $r_{\yA(\yB)}= \sqrt{1-|t_{\yA(\yB)}|^2}$, and $\chi=\frac{\w \D \ell}{v}+2\p\frac{\F}{\F_0}$ is the sum of the orbital and magnetic phase differences between electrons traversing the upper and lower MZI's arms with a geometric mismatch $\D \ell$.

The electronic MZI is coupled, through electrostatic interactions, to two detector QPCs (cf. QPC C and D in Fig.~\ref{fig:Setup}).
The electrons in the detectors are modeled in a similar fashion to the electrons in the MZI. It is useful to introduce the eigenmodes of QPC C:
\Eq{
	\vf_{mn}(x,\w;t)=\frac{e^{-i\w\bR{t-\frac{x}{v}}}}{\sqrt{L}}\condf{\I_{mn}&,x<x_\yC \\\bS{\cS_\yC(\w)}_{mn} & ,x>x_\yC}, \label{eq:QPCmodes}
}
where $m,n=3,4$. The eigenmodes of QPC D are defined similarly, replacing C with D and setting $m,n=5,6$. For simplicity of notations, throughout we assume that QPCs C and D are identical.
Electrostatic interactions between charges in the MZI arms and the detectors are modelled assuming that the presence of charge in the respective MZI arms slightly modifies the transmission probability of QPC C or QPC D, with strength proportional to $\lm$. The detectors sense charge fluctuations over a segment of length  $\ell$ in the respective interferometer arms.
For simplicity, we assume equal Fermi velocities $v$, and lengths $L$ of all channels, yielding the time-of-flight $\ta_{\rm FL}=L/v$.
The sources S$1$, S$3$ and S$5$ are biased by voltage $V$ relative to the other grounded contacts.

Measurements of zero-frequency cross-correlations and expectation values of the currents rely on readouts at the drains D$1$, D$3$, and D$5$ with the respective currents $\cI_{\yD 1}$, $\cI_{\yD 3}$, and $\cI_{\yD 5}$.
We now define a \ti{weak-weak-strong} (WWS) value of $\cI_{\yD 1}$. This employs the weakly measured signals  $\cI_{\yD 3}$ and $\cI_{\yD 5}$, and is given by
\EqS{
&\av{\cI_{\yD 1}}_{\rm WWS}\eqv\rav{\cI_{\yD 1}}_{\yD 3,\yD 5}-\rav{\cI_{\yD 1}}_{\yD 1,\yD 1},
\label{eq:TheSignal}
}
where $\rav{\cI_{\yD1}}_{\yD3,\yD5}\eqv\frac{\av{\dl \cI_{\yD1}\dl \cI_{\yD3}\dl \cI_{\yD5}}}{\av{\dl \cI_{\yD3}\dl \cI_{\yD5}}}$, and $\rav{\cI_{\yD1}}_{\yD1,\yD1}\eqv\av{(\dl \cI_{\yD1})^3}/\av{(\dl \cI_{\yD1})^2}$.
Here $\dl \cI\eqv \cI-\av{\cI}$ denotes the fluctuations of the current around its average value $\av{\cI}$.
The expectation values represent the low-frequency component of the signal and are obtained by averaging over a time window $\ta$, which we assume to be larger than all characteristic timescales of the experiment ($\ta\gg \ta_{\rm FL},\hb/eV$). For example, the three-current correlator is defined as
\EqS{
&\av{\dl \cI_{\yD1}\dl \cI_{\yD3}\dl \cI_{\yD5}} \equiv\\
&\equiv\lim_{\substack{\w_1\to0\\\w_2\to 0}}\frac{1}{\ta^2}\iint_{-\ta/2}^{\ta/2}dt_1dt_2 e^{i\w_1t_1} e^{i\w_2t_2}\cG_{135}(t_1,t_2),
\label{eq:ZeroFreqCorr}
}
where $\cG_{135}(t_1,t_2) \equiv \av{\dl \cI_{\yD1}(0)\dl \cI_{\yD3}(t_1)\dl \cI_{\yD5}(t_2)}$.

\paragraph{Single-particle analysis ---}

To lay out the concept, we first analyze a simplified single-particle picture. Physically, this corresponds to a large voltage/dilute current scenario, where the distance between consecutive electron wavepackets is larger than their spatial width \footnote{Here we neglect electron-electron interactions. This assumption is supported by the fact that experiments employing quantum Hall edges give rise to shot noise with Fano factor equals to 1, compatible with non-interacting electrons.}.
We also assume energy-independent transmission amplitudes, i.e., $\cA_{mn}(x,\w)\eqv\cA_{mn}(x)$.
In the absence of the detectors, the current at drain $\yD m$, $m=1,2$ which originates from a voltage biased source $\yS n$, $n=1,2$ is obtained via the Landauer-B\"uttiker formalism \cite{Landauer1957,Buttiker1990}, $\av{\cI_{\yD m}}= \frac{e^2 V}{h}|\mathcal{A}_{mn}(x_{\yD m})|^2$.
The current can be expressed in terms of the scattering matrices $\cS_\yB$, $\cS_\yA$ as
$\av{\cI_{\yD m}}=\braoket{\yS n}{\cI_{\yD m}}{\yS n}$, where $\cI_{\yD m}=\frac{e^2 V}{h}\cS_\yA\dg\cS_\yB\dg\cP_{\yD m}\cS_\yB\cS_\yA$. Here  $\cP_{\yD m}=\ketbra{\yD m}{\yD m}$ and $\ket{\yS n}$, $\ket{\yD m}$ are vector states (cf. the definition of $\cS_\yA$ and $\cS_\yB$) corresponding to the electron at source $\yS n$ and drain $\yD m$ respectively.

We unfold our Hilbert space in the system (MZI-detectors product space). Specifically, we consider  the propagation of a wavepacket
incident from the source $\yS1$, along with electrons in $\yS3$ and $\yS5$.
The incident state is thus
$\vert \Psi \rangle = \vert \yS1,\yS3,\yS5 \rangle$. The system-detector interaction is described by a scattering matrix of the QPC C, $\cS_{\yC}=\mat{r_d &-t_d\as\\ t_d&r_d}_{\yC}\otimes\I_{\yD}$. The transmission amplitude of QPC C depends on the position of the wave-packet in the MZI, i.e., $t_d=\tilde t_d\cdot\I_{\rm MZI}+\lm \cP_\yC$, where $\tilde t_d$ is the transmission in the absence of the wave-packet and $r_d=\sqrt{1-|t_d|^2}$.
Here $\lm$ is a parameter controlling the strength of the interaction and $\cP_\yC=\mat{1&0\\0&0}_{\rm MZI}$ is the projector to the upper MZI's arm.
Note that each element of $\cS_\yC$ is a $2\times 2$ matrix in the Hilbert space of the MZI's arms so that $\cS_{\yC}$ describes both the effect of the system on the detector signal and the back-action onto the system.
Similar considerations apply for QPC D, replacing the index C with D, and employing $\cP_\yD=\mat{0&0\\0&1}_{\rm MZI}$.
Analogously to the current at $\yD1$, the currents at $\yD3$ and $\yD5$ are defined by the expectation values of the matrices
$\cI_{\yD3}=\frac{e^2 V}{h}\cS_\yC\dg\cP_{\yD3}\cS_\yC$, and $\cI_{\yD5}=\frac{e^2 V}{h}\cS_\yD\dg\cP_{\yD5}\cS_\yD$ on the injected state $\ket{\Y}$, with $\cP_{\yD m}$  projectors on the arm $\yD m$.

We are now in a position to evaluate the WWS value of Eq.~\eqref{eq:TheSignal} within the single-particle framework,  $\av{\cI_{\yD1}}_{\rm WWS}^{\rm SP}$.
To compute this quantity we need the three-current correlator of Eq.~\eqref{eq:TheSignal}, which requires the ordering of the scattering matrices and the projectors along the wavelets' paths; these act from the sources $\yS1,\yS3,\yS5$ towards the drains $\yD1,\yD3,\yD5$ and backwards,
\EqS{
	&\av{\cI_{\yD1}\cI_{\yD3}\cI_{\yD5}}=\bR{\frac{e^2V}{h}}^3\times\\
	&\times\langle \Psi \vert \cS_\yA\dg \cS_\yC\dg\cS_\yD\dg\cS_\yB\dg\cP_{\yD1}\cP_{\yD3}\cP_{\yD5}\cS_\yB \cS_\yD\cS_\yC\cS_\yA \vert \Psi \rangle.
	\label{eq:SPFullCorrelator}
}

Likewise, in order to obtain an explicit form of  $\av{\cI_{\yD1}}_{\rm WWS}^{\rm SP}$ we need to compute the correlators $\av{\cI_{\yD3}\cI_{\yD5}}$, $\av{(\cI_{\yD1})^3}$, $\av{(\cI_{\yD1})^2}$ in terms of the scattering matrices $\cS_\yB$, $\cS_\yD$, $\cS_\yC$, $\cS_\yA$
and substitute in Eq.~\eqref{eq:TheSignal} (see Supplemental Material \footnote{see Supplemental Material.}).
The expression can be simplified considerably assuming $ t_d$ and $\lm$ to be real, leading to
\EqS{
\av{\cI_{\yD1}}_{\rm WWS}^{\rm SP}=&-\frac{\av{\com{\com{\cI_{\yD1}}{Q_\yC}}{Q_\yD}}}{4\av{Q_\yC}\av{Q_\yD}}.
\label{eq:SPSignal}
}
Here $Q_{\yC(\yD)}=\frac{e^2 V}{h}\frac{\ell}{v}\cS_\yA\dg \cP_{\yC(\yD)} \cS_\yA$ is an operator measuring charge sensed by the upper (lower) detector. We employ the explicit expressions of the charge and current operators to rewrite Eq.~\eqref{eq:SPSignal} as

\EqS{
	\av{\cI_{\yD1}}_{\rm WWS}^{\rm SP}=&-\frac{e^2 V}{h}\frac{\re{e^{i\chi}t_\yA t_\yB\as r_\yA r_\yB}
	}{2|t_\yA r_\yA|^2}.
	\label{eq:SPSignalExplicit}
}

\paragraph{Many-particle analysis ---}
The above single-particle analysis can be generalized to include a scenario where many particles are present and detected in the interferometer's arms. Throughout the following, we still discard electron-electron interaction within the MZI and within the detectors, yet account for the detection process (comprising interaction between a detector's electron and a MZI electron). The most important facet we want to include by accounting for such many-particle physics is that signals detected by the detectors and at the MZI drains may refer to different electrons (as opposed to partial waves of the same injected electron). Our formalism needs to rid of such spurious contributions.
Departing from a single-particle framework, we replace the Landauer-Bu\"ttiker approach by full-fledged time-dependent operator averages in Eq.~\eqref{eq:TheSignal}, evaluated within the Keldysh formalism. The three-current correlator (computed in the interaction picture, with the MZI
and the detectors being uncoupled) reads
\EqS{
	&\av{\hat \cI_{\yD1}(0)\hat \cI_{\yD3}(t_1)\hat \cI_{\yD5}(t_2)}=\\
	&=\av{\cT_\yK e^{-\frac{i}{\hb}\oint\hat \cH_{\rm MD}(t')dt'}\hat I_{\yD1}(0)\hat I_{\yD3}(t_1)\hat I_{\yD5}(t_2)}.
	\label{eq:MBCorrelator}
}
Here $\cT_\yK$
is the time-ordering operation  (along the Keldysh
time-contour) acting on the  Keldysh-symmetrized current operators
in the interaction picture, $\hat I_{\yD m}(t)=\hat U_0\dg(t) \hat\cI_{\yD m}(0) \hat U_0(t)$, $m=1,3,5$, where $\hat U_0(t)$ is the evolution operator with respect to the Hamiltonian of uncoupled MZI and detectors.

Quantum and thermal averaging is performed with respect to the density matrix, $\hat\vr(-\infty)$, describing the state of the impinging electrons (emitted from the  (possibly finite temperature) voltage biased reservoir), and the decoupled detectors C and D: $\hat \vr(-\infty)=\hat \vr_{\rm MZI}(-\infty)\otimes\hat \vr_{\text{QPC C}}(-\infty)\otimes \hat \vr_{\text{QPC D}}(-\infty)$.
The density matrix of the isolated MZI is expressed as
\begin{equation}
\hat \vr_{\rm MZI}(-\infty)=\prod_{n,\w} \bS{f_n(\omega)\hat{c}_n^{\dagger} (\omega) \hat{c}_n (\omega)+\bar f_n(\omega)\hat{c}_n (\omega) \hat{c}\dg_n (\omega)},
\label{eq:density}
\end{equation}
where $\hat c_n\dg(\w)$, $n=1,2$ is an operator creating an electron in the state $\psi_{mn}(x,\w;t)$ [Eq.~(\ref{eq:MZImodes})], $f_n(\omega)=\bR{1+e^{(\hb\omega-\mu_n)/k_B T}}\inv$ is the Fermi distribution of the electrons injected at $\yS n$, and $\bar f_n(\w)\eqv 1-f_n(\w)$. The density matrices of the detectors have analogous expressions with $c_n\dg(\w)$, $n=3,4,5,6$ defined through Eq.~(\ref{eq:QPCmodes}).

The current operator near $\yD1$ at time $t$ reads $\hat I_{\yD1}(t)=v \hat\ro_1(x_{\yD1};t)$, where the density operator $\hat\ro_m(x;t)=\hat \y_m\dg(x;t)\hat \y_m (x;t)$ and $\hat \y_m\dg(x;t)=\frac{\ta_{\rm FL}}{2\p}\int d\w\y_{mn}(x,\w;t)\hat c\dg_{n}(\w)$ is the operator that creates an electron in the $m$-th arm, at the position $x$ and time $t$. Throughout, we implicitly sum over repeated indices.
In order to express \eqref{eq:ZeroFreqCorr} in the frequency domain, and given the time averaging in \eqref{eq:ZeroFreqCorr}, all operators
therein should be evaluated at the same frequency, $\w$. Coherent superpositions of different frequency
components can then be ignored, which allows us to use
$\hat\ro_m(x;t)=\frac{\tau_{\rm FL}}{2\p}\int d\w [\ro_m(x,\w;t)]_{nl}\hat c\dg_{n}(\w) \hat c_{l}(\w)$
where $[\ro_m(x,\w;t)]_{nl}=e\y_{mn}(x,\w;t)\y_{ml}\as(x,\w;t)$. In turn, we are able to express the current operator as
\Eq{
	\hat I_{\yD1}(t)=\frac{h}{e V} \int \frac{d\w}{2\p}[\cI_{\yD1}(\w)]_{mn}\hat c_m\dg(\w)\hat c_n(\w).
\label{eq:corrente-operatore}
}
Analogous expressions hold for the detectors' currents $\hat I_{\yD3}(t)$, $\hat I_{\yD5}(t)$.
The matrices $\cI_{\yD m}(\w)$ are frequency-dependent generalizations of the matrices $\cI_{\yD m}$ appearing in the single-particle expressions, e.g., Eq.~\eqref{eq:SPFullCorrelator}.

The coupling between the MZI and the detectors is expressed through the Hamiltonian
\Eq{
\hat \cH_{\rm MD}(t)= \frac{\hb}{e^2}\bR{\tilde\lm\hat \G_{\yC}(t)\hat Q_\yC(t)+\tilde\lm \hat \G_{\yD}(t)\hat Q_\yD(t)}.
\label{eq:interazione}}
Here the charge and tunneling current operators are $\hat Q_\yC(t)=\int_{x\in \ell} dx \hat\ro_1(x;t)$, and $\hat \G_{\yC}(t)=\frac{\ta_{\rm FL}}{2\p}\int d\w [\tilde\G_\yC(\w;t)]_{mn}\hat c_{m}\dg(\w)\hat c_n(\w)$ respectively, where $[\tilde\G_\yC(\w;t)]_{mn}=iev\vf_{3m}(x_\yC^-,\w;t)\vf_{4n}\as(x_\yC^+,\w;t)+\rm{h.c.}$. Analogous expressions hold for $\hat Q_\yD(t)$ and $\hat\G_\yD(t)$, upon $\yC \leftrightarrow \yD$.

\begin{figure}
	\centering
	\includegraphics[width=8.6cm]{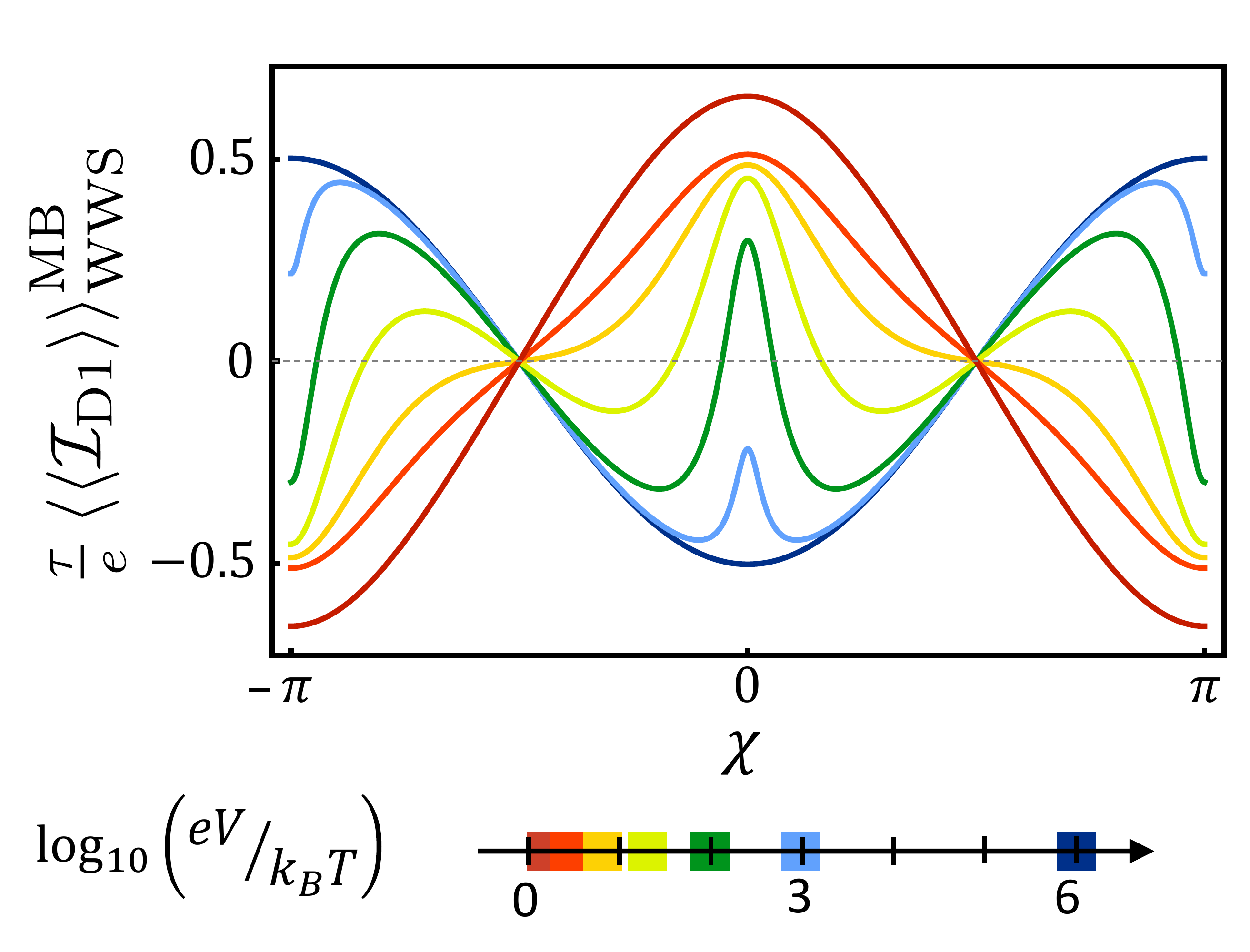}\\
	\caption[Cap1]{The WWS value, $\av{\cI_{\yD1}}_{\rm WWS}^{\rm MB}$ as function of the phase difference between the two arms, $\chi$, with transmission amplitudes $t_\yA=\sqrt{0.55}$ and $t_\yB=\sqrt{0.5}$, for several values of $eV/k_BT$. The WWS value oscillates around zero when varying $\ch$. In the two limiting cases, $eV/k_BT\gg 1$ and $eV/k_BT\ll 1$, the WWS value is proportional to $\av{\cI_{\yD1}}_{\rm WWS}^{\rm SP}$ [Eq.~\eqref{eq:SPSignalExplicit}], albeit with different proportionality coefficients. The latter reflect the temperature dependence of the two- and three-point correlation functions (see Eq.~(S20) in the Supplemental Material). \label{fig:MBSignal}}
\end{figure}

We evaluate the correlator in Eq.~\eqref{eq:MBCorrelator} to leading order in $\tilde\lm$ employing Eqs.~\eqref{eq:density}-\eqref{eq:interazione}. Similarly, we compute the other correlators in Eq.~\eqref{eq:TheSignal}. To obtain the WWS value in the many-body picture, $\av{\cI_{\yD1}}_{\rm WWS}^{\rm MB}$, we average the correlators over time according to Eq.~\eqref{eq:ZeroFreqCorr} \cite{Note3}.
 The results are presented in Fig.~\ref{fig:MBSignal} as a function of the interference control phase, $\chi$.

In the low-temperature regime, $eV\gg k_B T$, we find $\av{\cI_{\yD1}}_{\rm WWS}^{\rm MB}\at{eV\gg k_B T}=\frac{h}{\ta|eV|}\av{\cI_{\yD1}}_{\rm WWS}^{\rm SP}$. The r.h.s. represents the single-particle result weighted with the statistical probability of having a correlated noise, given that $\ta e|V|/h$ independent particles are injected during the measurement time, $\tau$. In the opposite, high-temperature limit, the signal converges to $\av{\cI_{\yD1}}_{\rm WWS}^{\rm MB}\at{eV\ll k_B T}=-\frac{2hk_BT}{|eV|^2}\av{\cI_{\yD1}}_{\rm WWS}^{\rm SP}$, with a prefactor reflecting the thermal noise and a reversed overall sign. The sign reversal arises from the change of the relative strengths of the two terms contributing to the WWS signal in Eq.~(\ref{eq:TheSignal}), following the analysis of the decoupled two- and three-point correlation functions \cite{Note3}. In both of these limits, $\av{\cI_{\yD1}}_{\rm WWS}^{\rm MB}$ oscillates around zero when varying $\chi$, but the oscillatory pattern changes non-trivially in the crossover from low to high temperature due many-body (thermal noise) effects.

\paragraph{Quantum vs. classical interference ---}
As demonstrated in Fig.~\ref{fig:MBSignal}, the three-point correlation function studied here assumes non-vanishing values for generic parameters of the interference setup.
In fact, a non zero value of $\av{\cI_{\yD1}}_{\rm WWS}$  is a direct signature of the quantum nature of the interference process.
In order to confirm this we show that our three-point correlator [Eq. (\ref{eq:TheSignal})] vanishes identically for two distinct classical scenarios: interference of classical waves, and (probabilistic, non-interfering) passage of classical particles through the interferometer arms. In both cases classical beam splitters replace the roles of the QPCs.

Consider first the case of classical particles. A particle emitted from $\yS1$ is scattered with probability $|t_\yA|^2$ onto arm $2$ and remains on the same arm with probability $|r_\yA|^2=1-|t_\yA|^2$.
The noisy detectors have the probability $|\tilde t_d|^2$ to click in the absence and probability $|\tilde t_d|^2+\lm$ to click in the presence of a particle in the upper (lower) arm.
To simplify the algebra, we set the rate of electrons injected at the detectors equal to the rate $W$ of electrons impinging at the beam splitter A.  We further assume that $W$ is small enough to have at most one particle in the segment between the beam splitters A and B at any instance of time.

The current correlations in this model can be determined following the same formalism outlined above (for the single-particle case), but replacing the coherent state $\vert \psi \rangle$ with a diagonal density matrix describing the probabilities of the classical particle to be in the respective interferometer arms.
 For the three-point correlator we obtain $\av{\dl \cI_{\yD1}\dl \cI_{\yD3}\dl \cI_{\yD5}}=-\lm^2 \cI_0^3 |t_\yA|^2 |r_\yA|^2 \bR{1-2\sI}$, where $\sI \equiv|t_\yA|^2 |t_\yB|^2+|r_\yA|^2 |r_\yB|^2$ and $\cI_0 \equiv eW$. The remaining two-point and self-correlators read $\av{\dl \cI_{\yD3}\dl \cI_{\yD5}}=-\lm^2 \cI_0^2 |t_\yA|^2 |r_\yA|^2$, $\av{(\dl\cI_{\yD1})^3}=\cI_0^3\sI(1-\sI)(1-2\sI)$, and $\av{(\dl\cI_{\yD1})^2}=\cI_0^2\sI(1-\sI)$. Finally, we obtain $\rav{\cI_{\yD1}}_{\yD1,\yD1}=\rav{\cI_{\yD1}}_{\yD3,\yD5}$, which, by Eq. (\ref{eq:TheSignal}), yields zero signal.

To establish a benchmark for classical waves, we consider a charge wave packet injected at S1 which, following the splitting at QPC A, propagates in the two arms $j=1,2$. The amplitude of the charge in arm $1$, $Q_{\yC}(t)$, is sensed by the corresponding detector, QPC C (cf. Fig.~\ref{fig:Setup}) via the detector's signal $I_{\textrm{D3}}(t) \propto |\tilde t_{d}|^2+\lm \frac{\ell}{v}Q_{\yC}(t-\ta_{\yC})/e+\xi_3(t)$, where $\ta_{\yC}$ is the time of flight from $\yS1$ to the point where the charge is detected and $\xi_3(t)$ is a stochastic noise at the detector. A similar expression holds for the detector's signal $I_{D5}(t)$ sensing the charge $Q_{\yD}$ in arm $2$ of the interferometer with an added noise $\xi_5(t)$. Importantly, there is no back-action here, hence the amplitudes $Q_{\yC}(t)$ and $Q_{\yD}(t)$ are unaffected by the measurement outcome (by the detectors' noise). As a result, if we assume that the wave injected at $\yS1$ has a stochastic component $\xi_1(t)$, the noise of the signal in $\yD1$ is unaffected by $\xi_3(t)$ and $\xi_5(t)$. Employing the fact that $\langle \xi_i(t) \xi_j(t') \rangle \propto \delta_{i,j}$, it follows that $\av{\cI_{\yD 1}}_{\rm WWS}=0$, reflecting the classical nature (no back-action) of this interferometry.

\paragraph{Conclusions ---}
We have constructed a protocol capable of addressing the ``quantumness'' of interference. The detection signal is non-zero in the case of quantum interference and vanishes for classical waves and for classical particles. We have addressed both the limit of (at most) a single particle present at the interferometer at any given moment, as well as the limit of many particles present. Our protocol does not require to register a signal as function of an externally varied parameter (e.g., the phase difference between the two arms).

Experimentally, to obtain the quantity represented by Eq.~\eqref{eq:TheSignal}, four different measurements are required:  the three-current cross-correlation of the currents in $\yD1$, $\yD3$ and $\yD5$ [cf.~Eq.\eqref{eq:SPFullCorrelator}], the two-current cross-correlation of the two detectors, and the two- and three- self-correlation functions of the current in two electronic beam splitters (QPC C and D). Note that for the electronic case both the interferometer and the QPCs operate in the quantum Hall regime, cf. Fig.~\ref{fig:Setup}. Given recent experimental advances in the field, we believe that our protocol can be implemented and verified in experiment. Furthermore, an intriguing follow up, both theory-wise and experiment-wise, would be the generalization of the above protocol to anyonic interferometry.

\paragraph{Acknowledgements ---} A.R. acknowledges support by EPSRC via Grant No. EP/P010180/1. Y.G. acknowledges funding from DFG RO 2247/11-1, CRC 183 (project C01), and the Minerva foundation.


%

\begin{widetext}
\newpage


\section*{Supplementary material}

# Detection of quantum interference without interference - Supplemental Material

## I. DERIVATION OF WEAK-WEAK-STRONG VALUE FOR THE SINGLE-PARTICLE CASE

In this section we compute the WWS value [Eq. (3)] employing a single-particle analysis. First, we express the noise correlations, defined below Eq. (3), through correlations of currents. This yields for the two- and three-noise correlators

$$\langle \delta \mathcal{I}_{D3} \delta \mathcal{I}_{D5} \rangle = \langle \mathcal{I}_{D3} \mathcal{I}_{D5} \rangle - \langle \mathcal{I}_{D3} \rangle \langle \mathcal{I}_{D5} \rangle, \quad (S1)$$

and

$$\langle \delta \mathcal{I}_{D1} \delta \mathcal{I}_{D3} \delta \mathcal{I}_{D5} \rangle = \langle \mathcal{I}_{D1} \mathcal{I}_{D3} \mathcal{I}_{D5} \rangle + 2\langle \mathcal{I}_{D1} \rangle \langle \mathcal{I}_{D3} \rangle \langle \mathcal{I}_{D5} \rangle - \langle \mathcal{I}_{D1} \rangle \langle \mathcal{I}_{D3} \mathcal{I}_{D5} \rangle - \langle \mathcal{I}_{D3} \rangle \langle \mathcal{I}_{D1} \mathcal{I}_{D5} \rangle - \langle \mathcal{I}_{D5} \rangle \langle \mathcal{I}_{D1} \mathcal{I}_{D3} \rangle. \quad (S2)$$

Similarly, the auto-correlators of $\mathcal{I}_{D1}$ are $\langle \delta \mathcal{I}_{D1}^2 \rangle = \langle \mathcal{I}_{D1}^2 \rangle - \langle \mathcal{I}_{D1} \rangle^2$, and $\langle \delta \mathcal{I}_{D1}^3 \rangle = \langle \mathcal{I}_{D1}^3 \rangle + 2\langle \mathcal{I}_{D1} \rangle^3 - 3\langle \mathcal{I}_{D1}^2 \rangle \langle \mathcal{I}_{D1} \rangle$.

In what follows, we evaluate the correlators in Eqs. (S1) and (S2) assuming weak MZI-detectors coupling. This allows us to expand all the expressions to leading order in the coupling strengths $\lambda_C$ and $\lambda_D$. First, we expand the scattering matrix of the QPC C, yielding

$$\mathcal{S}_C = \tilde{\mathcal{S}}_C \left( \mathbb{1} - i\lambda \left( \frac{h}{e^2 V} \right)^2 \frac{v}{\ell} \Gamma_C \mathcal{S}_A Q_C \mathcal{S}_A^\dagger \right) + \mathcal{O}(\lambda^2). \quad (S3)$$

Here $\mathcal{S}_C \equiv \tilde{\mathcal{S}}_C \big|_{\lambda_C=0}$, $Q_C = \frac{e^2 V}{h} \frac{\ell}{v} \mathcal{S}_A^\dagger \mathcal{P}_C \mathcal{S}_A$ is the charge operator within the segment $\ell$ detected by QPC C, and $\Gamma_C = \frac{e^2 V}{h} i \mathcal{S}_C^\dagger \frac{\partial \mathcal{S}_C}{\partial \tilde{t}_d} \big|_{\lambda=0}$ measures the tunneling current in QPC C due to a fluctuation in $Q_C$. Similar considerations apply for QPC D, replacing the index C with D.

Next, we evaluate the current-current correlators with respect to the injected wavefunction $|\Psi\rangle = |S1, S3, S5\rangle$. An expression for the generic correlator reads

$$\langle \mathcal{I}_{D1}^a \mathcal{I}_{D3}^b \mathcal{I}_{D5}^c \rangle = \left( \frac{e^2 V}{h} \right)^{a+b+c} \times \\ \times \langle \Psi | \mathcal{S}_A^\dagger \mathcal{S}_C^\dagger \mathcal{S}_D^\dagger \mathcal{S}_B^\dagger \mathcal{P}_{D1}^a \mathcal{P}_{D3}^b \mathcal{P}_{D5}^c \mathcal{S}_B \mathcal{S}_D \mathcal{S}_C \mathcal{S}_A | \Psi \rangle, \quad (S4)$$

for integer powers $a, b, c$. This notation covers all the correlators required for the computation of the WWS value (cf. Eqs. (S1) and (S2)).

We expand the correlators given in Eq. (S4) to lowest order in $\lambda$ employing Eq. (S3). The leading order of $\langle \delta \mathcal{I}_{D1} \delta \mathcal{I}_{D3} \delta \mathcal{I}_{D5} \rangle$ and $\langle \delta \mathcal{I}_{D3} \delta \mathcal{I}_{D5} \rangle$ is $\mathcal{O}(\lambda^2)$, whereas $\langle \delta \mathcal{I}_{D1}^2 \rangle$ and $\langle \delta \mathcal{I}_{D1}^3 \rangle$ have zero-order terms. Therefore, to this order, the ratio appearing in Eq. (3) is independent of the small parameter, $\lambda$, and reads

$$\langle\langle \mathcal{I}_{D1} \rangle\rangle_{D3,D5} = \\ = \frac{\mathfrak{Re} \left\{ \langle \delta Q_C \delta Q_D \delta \mathcal{I}_{D1} \rangle \mathcal{R}_C \mathcal{R}_D - \langle \delta Q_C \delta \mathcal{I}_{D1} \delta Q_D \rangle \mathcal{R}_C \mathcal{R}_D^* \right\}}{(\mathcal{R}_C - \mathcal{R}_C^*)(\mathcal{R}_D - \mathcal{R}_D^*)\langle \delta Q_C \delta Q_D \rangle}, \quad (S5)$$

where $\delta Q_{C(D)} = Q_{C(D)} - \langle Q_{C(D)} \rangle$, and $\mathcal{R}_{C(D)} \equiv \langle \Gamma_{C(D)}^\dagger \mathcal{I}_{D3(D5)} \rangle - \langle \Gamma_{C(D)}^\dagger \rangle \langle \mathcal{I}_{D3(D5)} \rangle$. All the expectation values in Eq. (S5) are calculated at $\lambda = 0$. An explicit form of $\mathcal{R}_C = \mathcal{R}_D$ reads $\mathcal{R}_C = i\mathfrak{Re}\{\tilde{t}_d\} - (1 - |\tilde{t}_d|^2)\mathfrak{Im}\{\tilde{t}_d\}$. We will assume $\mathfrak{Im}\{\tilde{t}_d\} = 0$, yielding $\frac{\mathcal{R}_C}{\mathcal{R}_C - \mathcal{R}_C^*} = \frac{1}{2}$. With this assumption Eq. (S5) reads

$$\langle\langle \mathcal{I}_{D1} \rangle\rangle_{D3,D5} = \frac{\overline{\langle \delta Q_C \delta Q_D \delta \mathcal{I}_{D1} \rangle}}{\overline{\langle \delta Q_C \delta Q_D \rangle}}, \quad (S6)$$

where

$$\overline{\langle \delta \mathcal{I}_{D1} \delta Q_C \delta Q_D \rangle} = \frac{1}{4} \big[ \langle \delta Q_C \delta \mathcal{I}_{D1} \delta Q_D \rangle + \langle \delta Q_D \delta \mathcal{I}_{D1} \delta Q_C \rangle \big] + \\ + \frac{1}{8} \big[ \langle \delta Q_C \delta Q_D \delta \mathcal{I}_{D1} \rangle + \langle \delta Q_D \delta Q_C \delta \mathcal{I}_{D1} \rangle + \\ + \langle \delta \mathcal{I}_{D1} \delta Q_C \delta Q_D \rangle + \langle \delta \mathcal{I}_{D1} \delta Q_D \delta Q_C \rangle \big], \quad (S7)$$

and

$$\overline{\langle \delta Q_C \delta Q_D \rangle} = \frac{1}{2} [\langle \delta Q_C \delta Q_D \rangle + \langle \delta Q_D \delta Q_C \rangle]. \quad (S8)$$

In the next section we derive Eq. (6) employing Eq. (S5).

## II. DERIVATION OF EQ. (6)

In this section we derive Eq. (6). For simplicity, we introduce dimensionless single particle charge and current operators, respectively defined as $\bar{Q}_{C(D)} = \mathcal{S}_A^\dagger \mathcal{P}_{C(D)} \mathcal{S}_A$, and $\bar{\mathcal{I}}_{D1(D2)} = \mathcal{S}_A^\dagger \mathcal{S}_B^\dagger \mathcal{P}_{C(D)} \mathcal{S}_B \mathcal{S}_A$. In the following derivation we will employ the current conservation and the non-locality of measurements on opposite arms, yielding:

$$\bar{Q}_C + \bar{Q}_D = \bar{\mathcal{I}}_{D1} + \bar{\mathcal{I}}_{D2} = \mathbb{1}_{MZI} \quad (S9a)$$
$$\bar{Q}_C \bar{Q}_D = \bar{\mathcal{I}}_{D1} \bar{\mathcal{I}}_{D2} = 0. \quad (S9b)$$

First, we compute $\overline{\langle \bar{Q}_C \bar{Q}_D \bar{\mathcal{I}}_{D1} \rangle}$, defined in Eq. (S7). It consists of a sum of three-noise correlators of $\bar{Q}_C$, $\bar{Q}_D$ and $\bar{\mathcal{I}}_{D1}$. Expanding the noise correlators according to Eq. (S2), and using relations from Eq. (S9) we arrive at

$$\overline{\langle \delta \bar{\mathcal{I}}_{D1} \delta \bar{Q}_C \delta \bar{Q}_D \rangle} = \frac{1}{4} \langle \big[ [\bar{\mathcal{I}}_{D1}, \bar{Q}_C], \bar{Q}_D \big] \rangle + \\ + 2\langle \bar{\mathcal{I}}_{D1} \rangle \langle \bar{Q}_C \rangle \langle \bar{Q}_D \rangle - \langle \bar{Q}_C \bar{\mathcal{I}}_{D1} \bar{Q}_C \rangle \langle \bar{Q}_D \rangle - \langle \bar{Q}_D \bar{\mathcal{I}}_{D1} \bar{Q}_D \rangle \langle \bar{Q}_C \rangle. \quad (S10)$$

Next, we compute the denominator of $\langle\langle \mathcal{I}_{D1} \rangle\rangle_{D3,D5}$, yielding

$$\overline{\langle \delta \bar{Q}_C \delta \bar{Q}_D \rangle} = -\langle \bar{Q}_C \rangle \langle \bar{Q}_D \rangle. \quad (S11)$$

To compute the WWS value (see Eq. (3)) we need to compute $\langle\langle\mathcal{I}_{D1}\rangle\rangle_{D1,D1}$ defined from the two- and three-noise correlators of $\bar{\mathcal{I}}_{D1}$. A simple calculation yields: $\langle(\delta\bar{\mathcal{I}}_{D1})^2\rangle = \langle\bar{\mathcal{I}}_{D1}\rangle(1-\langle\bar{\mathcal{I}}_{D1}\rangle)$, and $\langle(\delta\bar{\mathcal{I}}_{D1})^3\rangle = \langle\bar{\mathcal{I}}_{D1}\rangle(1-\langle\bar{\mathcal{I}}_{D1}\rangle)(1-2\langle\bar{\mathcal{I}}_{D1}\rangle)$. Combining the above results we obtain

$$\langle\mathcal{I}_{D1}\rangle^{SP}_{WWS} = \frac{\langle\overline{\delta\mathcal{I}_{D1}\delta Q_C \delta Q_D}\rangle + \langle\bar{Q}_C\rangle\langle\bar{Q}_D\rangle(1-2\langle\bar{\mathcal{I}}_{D1}\rangle)}{-\langle\bar{Q}_C\rangle\langle\bar{Q}_D\rangle}. \quad (S12)$$

To obtain Eq. (6), we need to show that

$$\langle\bar{Q}_C\rangle\langle\bar{Q}_D\rangle - \langle\bar{Q}_C\bar{\mathcal{I}}_{D1}\bar{Q}_C\rangle\langle\bar{Q}_D\rangle - \langle\bar{Q}_D\bar{\mathcal{I}}_{D1}\bar{Q}_D\rangle\langle\bar{Q}_C\rangle \quad (S13)$$

is zero. Employing Eqs. (S9) we rewrite the expression in Eq. (S13) in the form

$$\langle\bar{Q}_C\bar{\mathcal{I}}_{D2}\bar{Q}_C\rangle\langle\bar{Q}_D\bar{\mathcal{I}}_{D2}\bar{Q}_D\rangle - \langle\bar{Q}_C\bar{\mathcal{I}}_{D1}\bar{Q}_C\rangle\langle\bar{Q}_D\bar{\mathcal{I}}_{D1}\bar{Q}_D\rangle. \quad (S14)$$

In this form, the expression can be interpreted as the difference between the classical trajectories encircling the interferometer and going through D1 and D2. Such a difference is indeed zero.

## III. DERIVATION OF THE WEAK-WEAK-STRONG VALUE IN THE MANY-PARTICLE CASE.

Here we compute $\langle\mathcal{I}_{D1}\rangle_{WWS}$ via a many-body formalism. We employ the definition in Eq. (3), where we replace the current correlators by expectation values of the operators $\hat{\mathcal{I}}_{Dm}(t)$, $m=1,3,5$ measuring current in the drain D$m$ at time $t$, symmetrized along the Keldysh contour.

First, we compute the correlators making up $\langle\langle\mathcal{I}_{D1}\rangle\rangle_{D3,D5}$, averaged over time according to Eq. (4) to the order $\mathcal{O}(\tilde{\lambda}^2)$. To this end, we introduce the operators $\hat{I}_{Dm}(t)$ defined in the interaction picture with respect to the MZI-detectors interaction Hamiltonian [Eq. (11)]. This yields Eq. (8) for the three-current correlator. Similarly, the two-current correlator reads

$$\langle\hat{\mathcal{I}}_{D3}(0)\hat{\mathcal{I}}_{D5}(t)\rangle = \langle\mathcal{T}_K e^{-\frac{i}{\hbar}\oint \hat{\mathcal{H}}_{MD}(t')dt'} \hat{I}_{D3}(0)\hat{I}_{D5}(t)\rangle. \quad (S15)$$

Expanding these expressions to leading order in $\tilde{\lambda}^2$ and employing Eqs. (S1) and (S2), we arrive at the following expressions,

$$\langle\delta\hat{\mathcal{I}}_{D3}(0)\delta\hat{\mathcal{I}}_{D5}(t)\rangle = -\frac{\tilde{\lambda}^2}{e^4}\oint dt'_1 \oint dt'_2 \times \\ \times \mathcal{T}_K\langle\hat{\mathcal{R}}_C(t'_1,0)\rangle\langle\mathcal{R}_D(t'_2,t)\rangle\langle\delta\hat{Q}_C(t'_1)\delta\hat{Q}_D(t'_2)\rangle, \quad (S16)$$

and

$$\langle\delta\hat{\mathcal{I}}_{D1}(0)\delta\hat{\mathcal{I}}_{D3}(t_1)\delta\hat{\mathcal{I}}_{D5}(t_2)\rangle = -\frac{\tilde{\lambda}^2}{e^4}\oint dt'_1 \oint dt'_2 \times \\ \times \mathcal{T}_K\langle\hat{\mathcal{R}}_C(t'_1,t_1)\rangle\langle\hat{\mathcal{R}}_D(t'_2,t_2)\rangle\langle\delta\hat{I}_{D1}(0)\delta\hat{Q}_C(t'_1)\delta\hat{Q}_D(t'_2)\rangle, \quad (S17)$$

where $\hat{\mathcal{R}}_{C(D)}(t,t') \equiv \hat{\Gamma}_{C(D)}(t)\hat{I}_{D3(D5)}(t') - \langle\hat{\Gamma}_{C(D)}(t)\rangle\langle\hat{I}_{D3(D5)}(t')\rangle$. The Keldysh-contour ordering operator, $\mathcal{T}_K$, acts on all the operators to its right, yielding six different orderings which are illustrated in Fig. S1. Notice that there are additional terms arising due to different detection times in the MZI and the QPCs, which are suppressed by an averaging over a large time window. Summing over all these terms yields an expression similar to Eq. (S5), albeit with matrices replaced by many-body operators.

Now we are in a position to evaluate the correlation functions in Eqs. (S16) and (S17). To this end, we express each operator in these equations in terms of fermionic operators, $\hat{c}_n(\omega)$. For example, the charge density is defined as $\hat{\rho}_m(x;t) = \left(\frac{\tau_{FL}}{2\pi}\right)^2 \int d\omega d\omega' [\rho_m(x,t;\omega,\omega')]_{nl}\hat{c}_n^\dagger(\omega)\hat{c}_l(\omega')$ where $[\rho_m(x,t;\omega,\omega')]_{nl} = e\psi_{mn}(x,\omega;t)\psi^*_{ml}(x,\omega';t)$. Therefore, the charge and the current operators read, $\hat{Q}_{C(D)}(t) = \int_{x\in\ell_{C(D)}} dx\hat{\rho}_{1(2)}(x;t)$ and $\hat{I}_{D1}(t) = v\hat{\rho}_m(x_{D1};t)$. The tunneling current operator in QPC C reads $\hat{\Gamma}_C(t) = \left(\frac{\tau_{FL}}{2\pi}\right)^2 \int d\omega d\omega' [\tilde{\Gamma}_C(t;\omega,\omega')]_{mn}\hat{c}_m^\dagger(\omega)\hat{c}_n(\omega')$, respectively with $[\tilde{\Gamma}_C(t;\omega,\omega')]_{mn} = iev\alpha_C\varphi_{3m}(x_C^-,\omega;t)\varphi^*_{4n}(x_C^+,\omega';t) + $ h.c., and the current at D3 is defined as $\hat{I}_{D3}(t) = \left(\frac{\tau_{FL}}{2\pi}\right)^2 \int d\omega d\omega' [\mathcal{I}_{D3}(t;\omega,\omega')]_{mn}\hat{c}_m^\dagger(\omega)\hat{c}_n(\omega')$, where $[\mathcal{I}_{D3}(t;\omega,\omega')]_{mn} = \varphi_{3m}(x_{D3},\omega;t)\varphi^*_{3n}(x_{D3},\omega';t)$. Similar definitions apply for the operators in QPC D. For simplicity, we are assuming equal-arm MZI, $\Delta\ell=0$, and frequency independent tunneling coefficients at each of the QPCs.

Next, we perform the time integrals in each area in Fig. S1. This leads to a suppression of the non-diagonal-in-frequency expressions, giving rise to the single frequency integral definitions of the current and charge operators, appearing above Eq. (10) and below Eq. (11). Remarkably, the same many-body operators can be alternatively expressed via their single particle analogues, e.g., $\hat{I}_{Dm} = \frac{h}{eV}\int \frac{d\omega}{2\pi}[\mathcal{I}_{Dm}]_{nl}\hat{c}_n^\dagger(\omega)\hat{c}_l(\omega)$, $\hat{Q}_{C(D)} = \frac{h}{eV}\int \frac{d\omega}{2\pi}[Q_{C(D)}]_{mn}\hat{c}_m^\dagger(\omega)\hat{c}_n(\omega)$, and $\hat{\Gamma}_{C(D)} = \frac{h}{eV}\int \frac{d\omega}{2\pi}[\Gamma_{C(D)}]_{mn}\hat{c}_m^\dagger(\omega)\hat{c}_n(\omega)$.

Finally, we are left with time-independent two- and three- point correlators of the current and charge operators, containing a single frequency integral. To compute expectation values of the operators, we contract different fermionic operators employing the Wick's theorem, and their expectation values $\langle\hat{c}_n^\dagger(\omega)\hat{c}_n(\omega)\rangle = f_n(\omega)$, cf. Eq. (9). We thus derive general expressions for the correlators of two and three operators of the form, $\hat{O}_i = \frac{\tau_{FL}}{2\pi}\int d\omega[O_i]_{mn}\hat{c}_m^\dagger(\omega)\hat{c}_n(\omega)$ to arrive at,

$$\langle\delta\hat{O}_1\delta\hat{O}_2\rangle = \tau_{FL}\int\frac{d\omega}{2\pi}\text{Tr}\left\{O_1\bar{F}_\omega O_2 F_\omega\right\}, \quad (S18)$$

and

$$\langle\delta\hat{O}_1\delta\hat{O}_2\delta\hat{O}_3\rangle = \tau_{FL}\int\frac{d\omega}{2\pi}\times \\ \times \text{Tr}\left\{O_1\bar{F}_\omega O_2\bar{F}_\omega O_3 F_\omega - \bar{F}_\omega O_3 F_\omega O_2 O_1\right\}. \quad (S19)$$

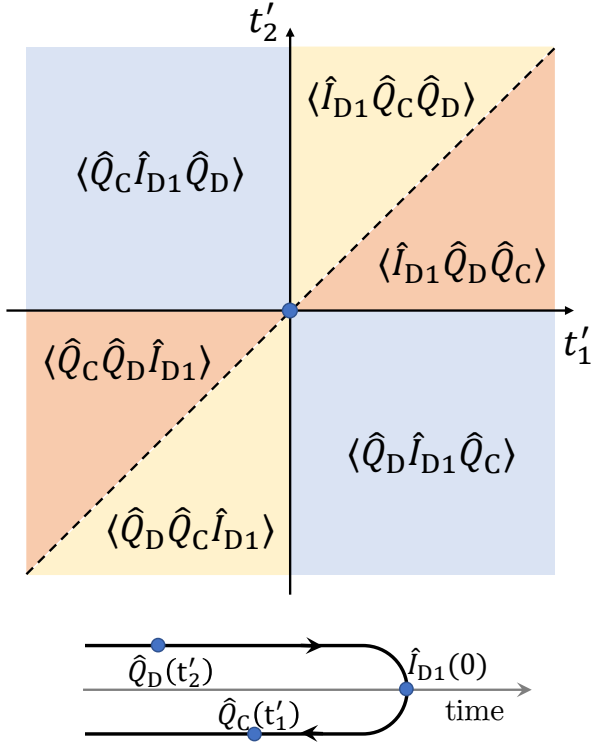

FIG. S1. Two-time integration over Keldysh contours. Upper panel: The area spanned by the two times, $t'_1$, at which $\hat{Q}_C$ is measured, and $t'_2$, at which $\hat{Q}_D$ is measured. The current $\hat{I}_{D1}$ is measured at $t = 0$. Negative (positive) times correspond to the negative (positive) forward (backward) Keldysh branch. The lines $t'_1 = 0$, $t'_2 = 0$ and $t'_1 = t'_2$ divide the plane into six regions, each corresponding to a different operator ordering. Lower panel: A demonstration of the operator ordering on the Keldsh time-contour.

Here $[F_\omega]_{mn} = f_m(\omega)\delta_{mn}$, and $\left[\bar{F}_\omega\right]_{mn} = \bar{f}_m(\omega)\delta_{mn}$.

To calculate $\langle\langle \mathcal{I}_{D1}\rangle\rangle_{D1,D1}$, we directly employ Eqs. (S18) and (S19), substituting $\hat{O}_1 = \hat{O}_2 = \hat{O}_3 \equiv \frac{h}{eV\tau_{FL}}\hat{\mathcal{I}}_{D1}$.

We consider the following correlators,

$$\langle(\delta\hat{I}_{D1})^2\rangle = \frac{e}{\tau}\langle\hat{I}_{D1}\rangle C_1(a) \quad \text{(S20a)}$$

$$\overline{\langle\delta\hat{Q}_C\delta\hat{Q}_D\rangle} = \langle\hat{I}_{D1}\rangle\frac{e\ell_C\ell_D}{\tau v^2}C_2(a) \quad \text{(S20b)}$$

$$\langle(\delta\hat{I}_{D1})^3\rangle = \frac{e^2}{\tau^2}\langle\hat{I}_{D1}\rangle C_3(a) \quad \text{(S20c)}$$

$$\overline{\langle\delta\hat{I}_{D1}\delta\hat{Q}_C\delta\hat{Q}_D\rangle} = \langle\hat{I}_{D1}\rangle\frac{e^2\ell_C\ell_D}{\tau^2 v^2}C_4(a). \quad \text{(S20d)}$$

To study the behavior of the correlators in the high temperature limit, we expand each $C_i(a)$ in the limit of $a \ll 1$. The lowest order of $a$ in the expansion is $a^{-1}$. The term of this order is proportional to $V^{-1}$. In the low voltage limit, its divergence in the correlator is canceled due to the multiplication by the average current $\langle\hat{I}_{D1}\rangle$, which is linearly proportional in $V$. Notice that orders lower than $a^{-1}$ in the expansion of $C_i(a)$ are not allowed, as they would lead to divergent correlators in the limit $V \to 0$.

We expect correlators consisting of an even number of current or charge operators to keep their sign under the transformation $V \to -V$, while odd correlators should reverse the sign under the same transformation. Therefore, $C_1(a)$ and $C_2(a)$ have only odd powers of $a$ in the expansion, i.e., $\sum_{n\geq 0} b_n a^{2n-1}$, while $C_3(a)$, and $C_4(a)$ have only even powers, $\sum_{n\geq 0} b_n a^{2n}$, where $b_n$ are coefficients dependent on the parameters of the system and are generically different for each $C_i(a)$. The four functions, $C_i(a)$, $i = 1,..4$ are demonstrated in Fig. S2.

We can predict the small $a$ expansion of $C_i(a)$ based on physical considerations. At high temperature, we expect the noise-current, $\langle(\delta\hat{I}_{D1})^2\rangle$ to be proportional to the Nyquist noise. Therefore, the leading order in the expansion of $C_1(a)$ is linear in $T$, i.e., is proportional to $a^{-1}$. The two-point correlator of the charges $\overline{\langle\delta\hat{Q}_C\delta\hat{Q}_D\rangle}$ is proportional to the product $\sim \langle\hat{Q}_C\rangle\langle\hat{Q}_D\rangle$. Therefore, we expect the product of two average charges to be quadratic in $V$, because each average charge is linear in $V$. As follows, the leading order of $C_2(a)$ is expected to be proportional to $a$. As the three-current self-correlator, $\langle(\delta\hat{I}_{D1})^3\rangle$, is expected to dependent on the temperature the leading order expected in $C_3(a)$ is $a^2$, leading to a decaying correlator $\sim 1/T^2$. Finally, we found that the leading order of $C_4(a)$ is a constant in $a$.

## IV. BUILDING BLOCK CORRELATORS FOR THE WWS VALUE.

In this section we investigate the two- and three-point correlators making up the WWS value (see Eq. (3)) at high temperature. We construct dimensionless quantities dividing the correlators by the average current, and constants, and plot them as a function of $a = eV/k_BT$.

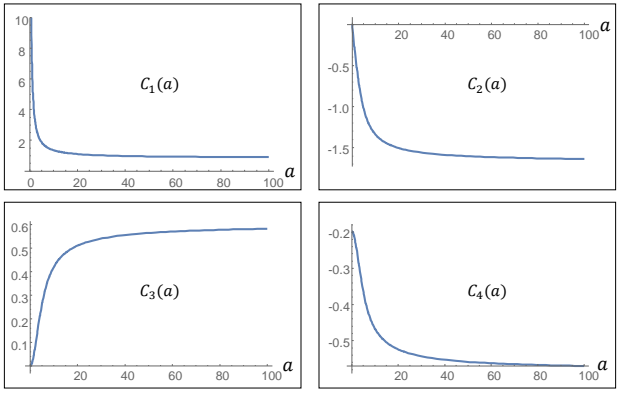

FIG. S2. Normalized correlators making up the WWS signal (see Eqs. (S20)), as a function of $a = eV/k_B T$, shown for $t_A = \sqrt{0.55}$, $t_B = \sqrt{0.5}$ and $\chi = \pi/4$.


\end{widetext}

\end{document}